\documentclass[12pt,preprint,twocolumn,hidepacs,superscriptaddress,twocolumn,
floatfix,preprintnumbers,altaffilletter]{aastex61}

\usepackage{apjfonts}
\usepackage{natbib}
\usepackage{psfig}
\usepackage{amsmath}
\def\be{\begin{eqnarray}}
\def\ee{\end{eqnarray}}
\usepackage{comment}
\usepackage{graphicx,subfigure}
\usepackage{appendix}
\usepackage{enumerate}
\shorttitle{The double-peaked radio light curve of PTF11qcj}
\shortauthors{Palliyaguru et al.}

\begin{document}
\title{The double-peaked radio light curve of supernova PTF11qcj}

\author{N.T.~Palliyaguru}\affiliation{Texas Tech University, Physics Department, Box 41051, Lubbock, TX 79409, USA}\affiliation{Arecibo Observatory, HC3 Box 53995, Arecibo, PR 00612, USA}
\author{A.~Corsi}\affiliation{Texas Tech University, Physics Department, Box 41051, Lubbock, TX 79409, USA}
\author{D.A.~Frail}\affiliation{National Radio Astronomy Observatory, 1003 Lopezville Road, Socorro, NM 87801, USA} 
\author{J.~Vink\'o}\affiliation{Konkoly Observatory, Research Centre for Astronomy and Earth Sciences, Hungarian Academy of Sciences, Konkoly-Thege M. \'ut 15-17, Budapest, 1121, Hungary}\affiliation{Department of Optics and Quantum Electronics, University of Szeged, D\'omt\'er 9, Szeged, 6720 Hungary}\affiliation{Department of Astronomy, University of Texas at Austin, Austin, TX, 78712, USA}
\author{J.C.~Wheeler}\affiliation{Department of Astronomy, University of Texas at Austin, Austin, TX, 78712, USA} 
\author{A.~Gal-Yam}\affiliation{Benoziyo Center for Astrophysics, Weizmann Institute of Science, 76100 Rehovot, Israel}
\author{S.B.~Cenko}\affiliation{Astrophysics Science Division, NASA Goddard Space Flight Center, Mail Code 661, Greenbelt, MD 20771, USA}\affiliation{Joint Space-Science Institute, University of Maryland, College Park, MD 20742, USA}
\author{S.R.~Kulkarni}\affiliation{Division of Physics, Math and Astronomy, California Institute of Technology, 1200 East California Boulevard, Pasadena, CA 91125, USA} 
 \author{M.M.~Kasliwal}\affiliation{Division of Physics, Math and Astronomy, California Institute of Technology, 1200 East California Boulevard, Pasadena, CA 91125, USA}


\begin{abstract}
We present continued radio follow--up observations of PTF11qcj, a highly energetic broad-lined Type Ic supernova (SN), with a radio peak luminosity comparable to that of the $\gamma$--ray burst (GRB) associated SN\,1998bw. The latest observations, carried out with the Karl G. Jansky Very Large Array (VLA), extend up to $\sim$5 years after the PTF11qcj optical discovery.
The radio light curve shows a double--peak profile, possibly associated with density variations in the circumstellar medium (CSM), or with the presence of an off-axis GRB jet.
Optical spectra of PTF11qcj taken during both peaks of the radio light curve do not show the broad H$\alpha$ features typically expected from H--rich circumstellar interaction. Modeling of the second radio peak within the CSM interaction scenario requires a flatter density profile and an enhanced progenitor mass--loss rate compared to those required to model the first peak. Although our radio data alone cannot rule out the alternative scenario of an off--axis GRB powering the second radio peak, the implied off-axis GRB parameters are unusual compared to typical values found for cosmological long GRBs. Deep X--ray observations carried out around the time of the second radio peak could have helped distinguish between the density variation and off-axis GRB scenarios. Future VLBA measurements of the PTF11qcj radio ejecta may unambiguously rule out the off-axis GRB jet scenario. 
\end{abstract}



\section{Introduction}
\label{intro}
The reason why some massive stars explode as supernovae (SNe) and others as rare $\gamma$--ray bursts (GRBs) remains a mystery.
Type Ib/c supernovae (SNe) are the result of the core collapse of massive stars, specifically the ones that have shed their hydrogen, and possibly helium, envelopes.
Massive Wolf Rayet (WR) stars and stars in close binary systems, that have completely lost their outer hydrogen layer due to stellar wind or through Roche-lobe overflow, respectively, are possible progenitors of these SNe \citep{ew88,g16}.
Long duration ($\gtrsim$ 2\,s) GRBs, with their engine--driven collimated outflows, are also thought to originate from the core collapse of massive stars \citep{wb06,p98}, being a rare subset of Type Ib/c SNe.
The leading scenario is that all long GRBs are accompanied by core--collapse SNe (though not all Ib/c SNe are accompanied by GRBs). In some long GRBs, the SN light may go undetected due to reasons such as large distances, poor localizations, dust extinction and galaxy contamination \citep{wb06}. It is noteworthy, however, that some nearby, well-localized GRBs without significant dust obscuration have been identified without a SN association \citep{fwt06,gfp06}.

So far,  $\sim 11$ Type Ib/c SNe have been discovered in connection with GRBs \citep{mlb16}, pointing to a relationship between the two events.
These include SN 1998bw \citep{gvp98} associated with GRB 980425, and a few more SNe \citep[2003dh, 2003lw, 2006aj, 2010bh, 2010ma, 2012bz, 2013cq, 2013dx, 2013ez;][]{smg03, mgs03, mtc04, pmm06, bps12, ssf11, smc14, xpl13, dpm15, cpp14}. SNe with a GRB association are generally more energetic than typical Type Ib/c SNe \citep{bkf03,mdt03,mdp06}, with explosion kinetic energies of $\sim10^{52}$ erg \citep{mmw14} for the former, and $\sim10^{51}$ erg \citep{tsl15} for the latter, and
also have broad features in their optical spectra \citep[BL-Ic;][]{wb06,g16} that imply high photospheric velocities.

In the radio, most "ordinary" BL-Ic SNe go undetected or are "radio quiet" \citep[$\rm L_{GHz}\lesssim 10^{26}$\,erg\,s$^{-1}$\,Hz$^{-1}$;][]{bkf03,sbk06,cgk16}. On the other hand, the GRB-associated SN\,1998bw was three orders of magnitude more radio luminous than, for example, the ordinary BL-Ic SN\,2002ap, although a few orders of magnitude less radio-luminous than cosmological GRB afterglows \citep[see e.g.,][and references therein]{kfw98,cf12, cgk16}.  Because radio emission probes the fastest ejecta, radio-loud BL-Ic SNe are more likely to be engine-driven (i.e., associated with GRBs). However, the radio-loud Ib/c SN\,2009bb showed no clear evidence for an association with a (high-luminosity) GRB \citep{scp10,pss11}, opening the question of whether there is  a class of core-collapse explosions with properties in between ordinary BL-Ic SNe and GRBs. In fact, the $\gamma$-ray energy of several GRBs with a spectroscopic SN association is also lower than that of typical cosmological GRBs \citep{aft02,mmw14}. This suggests that low-luminosity GRBs themselves may represent a distinct population of intrinsically lower-energy events \citep{bnp11,w04}, although (for at least some of them) an interpretation as ordinary GRBs observed off-axis is also possible \citep{yyn03,el99}.

Off-axis GRBs are a natural expectation of the fireball model \citep[e.g.][]{r97,p04}. An off-axis GRB accompanying a relativistic (engine-driven) BL-Ic SN should become visible at late times in the radio \citep{pl98,p01,w04}, thus representing a potential source of radio-loud emission with characteristic timescales ($\gtrsim 100$\,d since explosion) much longer than the radio peak time of 1998bw-like SNe \citep[10-20 d since explosion;][]{kfw98}.  The discovery of an off-axis long GRB associated with a BL-Ic SN remains, as of today, yet to be achieved.
In fact, all previous claims of off-axis GRB discoveries  \citep{gr04,ptk10} have subsequently been ruled out \citep{bb07,sbc10,bcg14}. 

One of the challenges in searching for off-axis events is the fact that the characteristic late-time peaking radio light curve of an off-axis GRB (whose emission enters our line of sight after substantial deceleration has occurred) may, at first glance, look similar to that of a non-relativistic SN whose ejecta are interacting strongly with a dense CSM. In radio SNe powered by synchrotron self-absorbed emission, the radio peak luminosity and the peak timescale probe the ejecta speed. Particularly, for a given radio luminosity, the later the peak time, the smaller the ejecta speed \citep{c98,bkf03}. A relatively slow turn--on in the radio has been observed in radio-bright non-relativistic Ib/c SNe such as SN\,1979C and SN\,1988Z \citep{wsp86,vws93}, thought to be  powered by shock interaction with a high-density medium \citep{vws93}. 
SN\,2001em, SN\,2003bg, SN\,2004cc, SN\,2004dk, SN\,2004gq, and SN\,2007bg \citep{wan03,gr04,sck06,qry07,wsc12,sbs13} are other more recent examples of (non-relativistic) SNe that exhibit late--time radio emission arising from CSM density variations. In summary, distinguishing between CSM-interaction and off-axis GRB jets requires an accurate analysis of broad-band datasets.

\begin{figure}
    \centering
\includegraphics[width=0.55\textwidth,trim={ 1.0cm 12.0cm 0cm 2.5cm},clip]{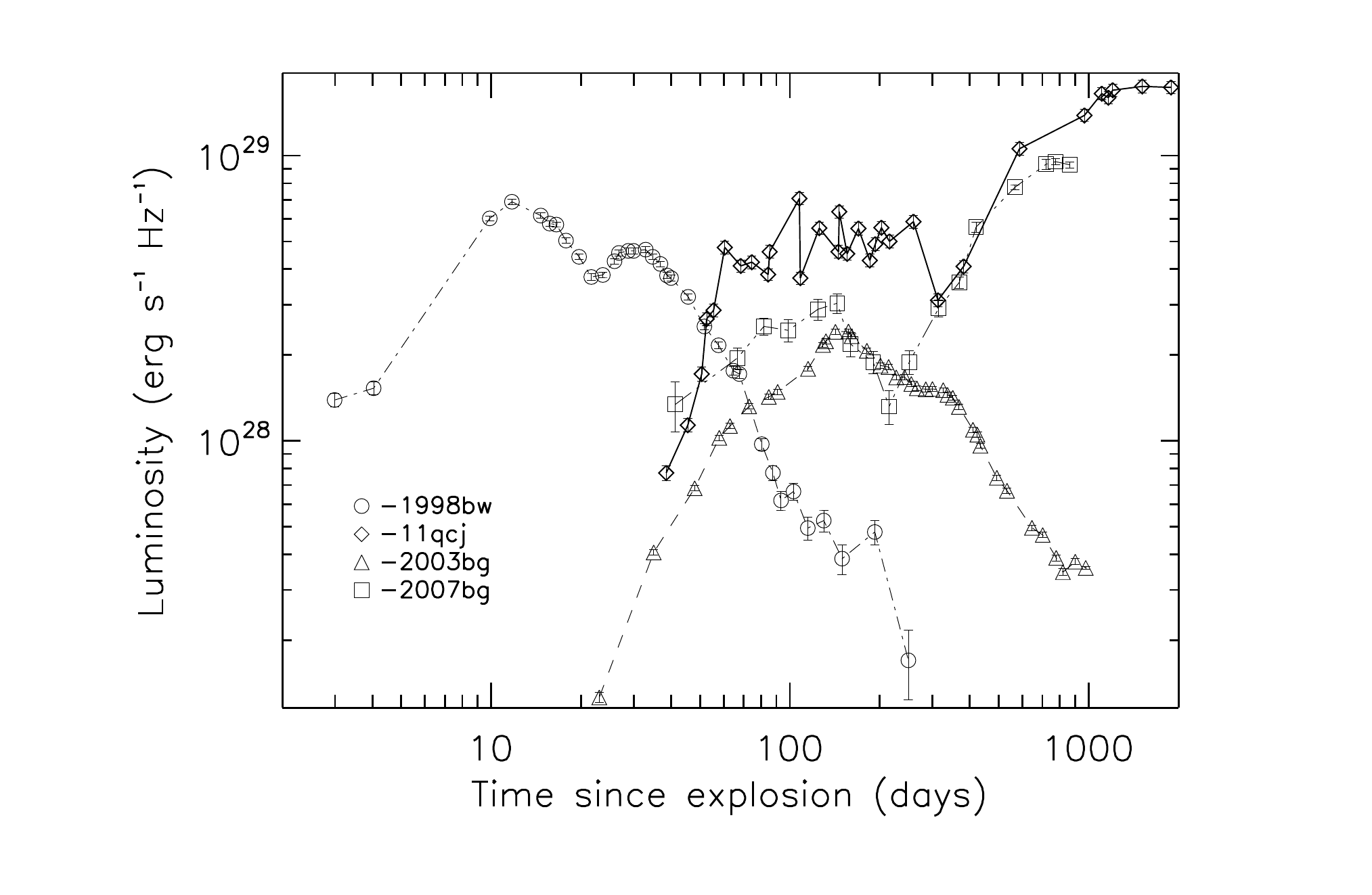}
\caption{
Radio luminosity of PTF11qcj compared to the GRB--associated SN 1998bw and other Ibc/IIb supernovae 2003bg \citep{skb05} and 2007bg \citep{sbs13} that show late-time radio re-brightening.
All data are at 5 GHz.}
\label{fig:lum_comp}
\end{figure}

Here, we present late-time radio observations of PTF11qcj, a bright BL-Ic SN for which the radio luminosity is $\sim 10^{29} \rm {erg\,s^{-1}\,Hz^{-1}}$, comparable to that of SN 1998bw \citep[see for e.g. Figure~\ref{fig:lum_comp} of this paper;][]{cog14}. Our extended radio follow-up observations of PTF11qcj show evidence for the presence of a second, late-time peak in its radio light curve. While the first radio peak pointed to a speed of the fastest ejecta of $\approx 0.3-0.5\,c$ and a high progenitor mass-loss rate ($\sim 10^{-4}$\,M$_{\odot}$\,yr$^{-1}$) indicative of strong CSM interaction  \citep{cog14}, here we focus on the analysis of the second late-time radio peak within the two possible scenarios of strong CSM interaction and off-axis GRB. Our extensive radio dataset is presented in Section~2; modeling of the radio data is discussed in Section~3; and results are presented in Section~4.
In Section~5 we conclude.

\section{Observations}
PTF11qcj was discovered by the Palomar Transient Factory \citep[PTF;][]{rkl09} on November 1st, 2011 UTC (MJD 55866), at $\alpha=13^{\rm{h}}13^{\rm{m}}41.51^{\rm{s}},\,\delta=+47^\circ 17\arcsec 57.0\arcsec$, and at a redshift of $z=0.028$, corresponding to a luminosity distance of $d_L\approx124$ Mpc \citep{cog14}.
Early optical observations and VLA follow--up data are presented in \citet{cog14}.
In this Section we describe later radio follow--up observations and spectroscopic data.

\subsection{Radio follow--up}
The VLA follow--up observations presented here were carried out between June 1st 2014 (MJD 56809) and December 7th 2016 (MJD 57729).
The data were taken at the nominal central frequencies of 2.5, 3.5, 7.4, 13.5 and 16 GHz with a nominal bandwidth of 2 GHz.
3C286 and J1327+4326 were used as flux and phase calibrators, respectively.

\begin{table}
\begin{center}\begin{footnotesize}
\caption{Summary of the late-time VLA observations of PTF11qcj. From left to right: observation start time (MJD), epoch in days since the estimated explosion date \citep[MJD 55842; see][]{cog14}, array configuration, central frequency, and flux density.
\label{tb:obs_info} }
\begin{tabular}{cclll}
\hline
\hline
Start time & Epoch & Observatory & Freq. & Flux Density
\\ (MJD)& (days)&  & (GHz)  & (mJy beam$^{-1}$)\\
\hline
56809.993&967 & VLA:A&2.5&4.01$\pm$0.20\\
"& "& VLA:A& 3.5& 6.10$\pm$0.30\\
"& "& VLA:A&5.0&7.55$\pm$0.37\\
"& "& VLA:A& 7.4&7.73$\pm$0.38\\
"& "& VLA:A&13&4.92$\pm$0.24\\
"& "& VLA:A& 16&4.09$\pm$0.20 \\
\hline
56948.910&1106  & VLA:C & 3.5& 6.74$\pm$0.33\\
"&  "& VLA:C & 5.0& 9.00$\pm$0.45\\
"&  "& VLA:C &7.4 &7.90$\pm$0.39 \\
"&  "& VLA:C & 13&5.50$\pm$0.27 \\
"&  "& VLA:C & 16& 4.66$\pm$0.23\\
\hline
57006.658&1164& VLA:C&2.5&6.18$\pm$0.30\\
"& "&VLA:C&3.5&7.00$\pm$0.35\\
"& "&VLA:C&5.0&8.71$\pm$0.43\\
"& "&VLA:C&7.4&8.39$\pm$0.41\\
"& "&VLA:C&13&5.76$\pm$0.28\\
"& "&VLA:C&16&4.90$\pm$0.24\\
\hline
57046.262&1204& VLA:CnB&5.0&9.25$\pm$0.46\\
"& "& VLA:CnB&7.4&9.49$\pm$0.47\\
\hline
57354.492&1512   & VLA:D&  13&    5.02$\pm$      0.25\\
"& "   & VLA:D&  16&     4.29$\pm$      0.21\\
"& "  & VLA:D&  5.0&      9.54$\pm$      0.47\\
"& "  & VLA:D&  7.3&      8.33$\pm$      0.41\\
"& "   & VLA:D&   2.4&      8.21$\pm$      0.41\\
"& "   & VLA:D&   3.4&     9.23$\pm$      0.46\\
\hline
57729.458&1887&VLA:A&       13&       3.62$\pm$      0.18\\
"&"&VLA:A&       16&       2.86$\pm$      0.14\\
"&"&VLA:A&       5.0&       9.46$\pm$      0.47\\
"&"&VLA:A&       7.3&       7.15$\pm$      0.35\\
"&"&VLA:A&       2.4&       7.80$\pm$      0.39\\
"&"& VLA:A&      3.4&       9.94$\pm$      0.49\\
\hline
\end{tabular}
\end{footnotesize}\end{center}\end{table}

The Common Astronomy Software applications \citep[CASA;][]{mws07} \footnote{Available online at http://casa.nrao.edu}
 was used to calibrate, flag and image the data.
The automated VLA calibration pipeline for CASA was used to calibrate the raw data.
Images were formed from the visibility data using the CLEAN algorithm \citep{h74}.
The image size was set to (1024$\times$1024) pixels, and the pixel size was determined as $1/4$ of the nominal beam width.
The images were cleaned using natural weighting for 10000 iterations or until a threshold of $\sim$0.03 mJy ($\sim3\sigma$) was reached.

The source flux was calculated as the flux corresponding to the brightest pixel
within a circle centered around the PTF position with a radius of $2\,\arcsec$ \citep[comparable to the typical $R$--band seeing of PTF images;][]{lkd09}.
Figure~\ref{fig:lc_all_models_phase2}
shows the radio light curves of PTF11qcj, with the new
 data from MJD 56809 onwards plotted along with the data from \citet{cog14}.
Flux errors are calculated as the quadratic sum of the rms map error and a 5\% fractional error that accounts for errors in the flux calibration \citep{wsp86,cog14}.
We note that results from \textit{imfit}, that fits a two dimensional elliptical Gaussian to a source, do not show evidence for extended emission, confirming that the radio counterpart of PTF11qcj is a point source up to $\approx 0.27$\,arcsec, which is the beam size in the VLA A array configuration at 7.4\,GHz.
The fluxes at each MJD and frequency for the latest observations are reported in Table~\ref{tb:obs_info}.

\subsection{Spectroscopic follow--up}
The location of PTF11qcj was observed with the Low Resolution Spectrograph-2 (LRS-2) attached to
the Hobby-Eberly Telescope (HET) on February 22, 2017 UT.
LRS-2 is a twin IFU spectrograph having
two arms each:  LRS2-B covers the range between 3700--4700 \AA\, (blue arm) and 4600--7000 \AA\,
(orange arm) with resolving power of 1900 and 1100, respectively, while LRS2-R covers the range from
6500 to 8400 \AA\, (red arm) and from 8200 to 10500 \AA\, (far-red arm) at a resolution of
1800 in each arm \citep{chonis14, chonis16}. Each IFU maps a 12" $\times$ 6" area on
the sky covered by 280 fibers. The coverage is complete, so no dithering is required.

\begin{figure}
    \centering
\includegraphics[width=0.45\textwidth,trim={ 1cm 0cm 2cm 17cm},clip]{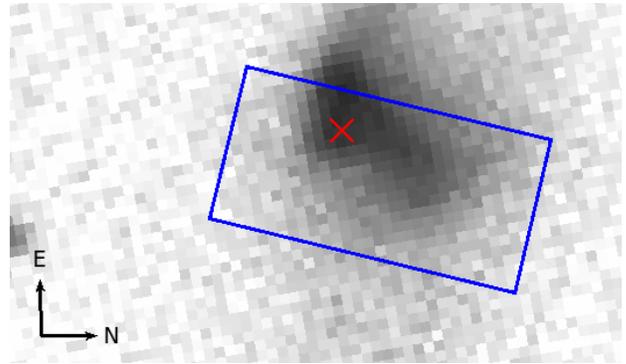}
\caption{
SDSS image of the host
galaxy of PTF11qcj. The SN position is marked with the
red cross. The position of the LRS2 IFU is indicated by the blue rectangle.
}
\label{fig:lrs2_sdss}
\end{figure}

We utilized the red arm of LRS2-R to collect spectra in the vicinity of the SN.
Figure~\ref{fig:lrs2_sdss} shows the SDSS $r$-band frame of the host galaxy of PTF11qcj with the
SN position marked in red. The blue rectangle indicates the position of the LRS2-R IFU (it was
not completely centered on the SN due to a minor pointing issue). A spectrum at the SN position was extracted by median-combining the signal of
the three closest fibers. Wavelength calibration was computed using FeAr spectral
lamp observations. Flux calibration was performed by comparing the
observed and flux-calibrated catalogued spectra of the standard star HD~84937.

\begin{figure}
    \centering
\includegraphics[width=0.52\textwidth,trim={ 1cm 1cm 6cm 17cm},clip]{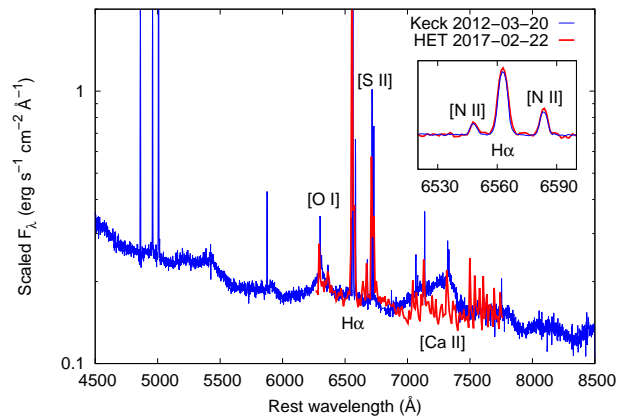}
\caption{HET-LRS2 spectrum with an exposure time of 2400\,s from 2017 February 22 UT taken at the position of PTF11qcj (red curve) and the Keck DEIMOS spectrum from 2012 March 20 UT (blue curve).
The inset zooms in on the narrow H$\alpha$ lines visible in the recent HET-LRS2 spectrum, which are nearly identical to those in the older Keck spectrum,  confirming that they are from the
host.
}
\label{fig:lrs2_sp}
\end{figure}

The final LRS2 spectrum of PTF11qcj is plotted in Figure~\ref{fig:lrs2_sp} together with the
spectrum obtained with the Keck2-DEIMOS on 2012-03-20 \citep{cog14}.
This plot suggests that the broad SN features that were clearly visible
in the Keck spectrum 5 years ago (typical nebular features due to forbidden transitions of
neutral oxygen [O I] $\lambda\lambda 6300,6364$ and ionized [Ca II] $\lambda\lambda 7291,7324$), may still be present in the new HET-LRS2 spectrum although with a smaller signal--to--noise.
Also, the bright, narrow H$\alpha$ feature present in both spectra appears basically
unchanged (in both strength and width). No broad H$\alpha$ feature can be identified in the HET-LRS2 spectrum of PTF11qcj (as was the case for the older Keck spectrum during the first radio peak), thus excluding H-rich CSM interaction similar to SN\,2014C \citep{mmk15}.  As discussed in what follows,  interaction with an
 H-poor CSM may explain the late-time radio re-brightening.

\section{Radio modeling}
As shown in  in Figure~\ref{fig:lc_all_models_phase2}, the higher-frequency (13.5 and 16 GHz) radio light curves of PTF11qcj appear double--peaked. At lower frequencies, a late-time re-brightening is also evident.
We thus identify two phases in the evolution of the radio emission, with the first peak in between $\sim$MJD 55842 and $\sim$MJD 56101, and the second starting from $\sim$MJD 56101.

As discussed in Section \ref{intro}, bright radio emission in a SN may be powered by the presence of a central engine, or via the interaction of the ejecta with the the CSM. Within the CSM-interaction scenario, a double-peaked radio light curve may be attributed to density variations in the CSM  \citep{sck06} due to variable (potentially eruptive) mass loss from the SN progenitor, or clumping due to turbulence in the medium and violent outbursts \citep{wsc12,sbs13}. In the case of a binary SN progenitor, variable radio emission may occur due to two different wind components from the two stars in the binary and/or the interaction of the shock with the common envelop or the wind termination shock \citep[see e.g.,][]{wsc12}. On the other hand, in the engine-driven scenario, a late--time peak in the radio light curve may be expected from an off--axis GRB: as the ultra-relativistic jet transitions to the sub-relativistic regime, it also spreads sideways resulting in a re-brightening of the radio SN light curve \citep[e.g.,][]{w04}.

In light of the above, in what follows we model the PTF11qcj light curves within both the synchrotron self-absorption (SSA) model with CSM density variations \citep{skb05}, and the off-axis GRB afterglow model \citep{em11,ehm12}.

\subsection{CSM-interaction SSA model}
\citet{skb05} model describes the radio emission from SN ejecta interacting with the CSM.
The radio flux density at time $t$ and frequency $\nu$ is given by:
\begin{equation}
F\left(t,\nu\right)=C_f\left(\frac{t-t_e}{t_0}\right)^{(4\alpha_r-\alpha_B)/2}\left[1-\rm {exp}^{-\tau_\nu^{\xi}(t)}\right]^{1/\xi}\nu^{5/2}F_3(x)F_2^{-1}(x), 
\end{equation}
with the optical depth given by
\begin{equation}
\tau_\nu\left(t\right)=C_\tau\left(\frac{t-t_e}{t_0}\right)^{(3+p/2)\alpha_B+(2p-3)\alpha_r-2(p-2)},
\end{equation}
where $C_f$, $C_\tau$ are normalization constants, $F_2$, $F_1$ are Bessel functions and $x=2/3(\nu/\nu_m)$ where $\nu_m$ is the critical synchrotron frequency, $t_0$ is a reference epoch, $t_{e}$ is the explosion time, $p$ is the electron energy index, $\xi=[0,1]$ describes the sharpness of the spectral break between optically thin and thick regimes \citep{skb05}.
In the above equations, $\alpha_r$ and $\alpha_B$ are the temporal indices of the shock radius $r$ and the magnetic field $B$ respectively, such that
\begin{equation}
r=r_{0}\left(\frac{t-t_e}{t_{0}}\right)^{\alpha_r},
\end{equation}
and
\begin{equation}
B=B_{0}\left(\frac{t-t_e}{t_{0}}\right)^{\alpha_B}, 
\end{equation}
where $r_0$ and $B_0$ the radius and magnetic field at the reference epoch $t_0$.
The expansion of the SN shock is described by: 
\begin{equation}
\alpha_r=\frac{n-3}{n-s},
\end{equation}
where $\rho_{ej}\propto r^{-n}$ is the density profile of the outer SN ejecta, and
 $\rho_{CSM}\propto n_e \propto r^{-s}$ that of the shocked CSM (or shocked electron density).
The self--similar conditions $s<3$ and $n>5$ \citep{c82} result in $\alpha_r<1$. 

In the standard scenario, the magnetic energy density and the relativistic
electron energy density are constant fractions, $\epsilon_B$ and $\epsilon_e$ respectively, of the post-shock energy density. Under these assumptions:
\begin{equation}
\alpha_B=\frac{(2-s)}{2}\alpha_r-1,
\end{equation}
and the minimum Lorentz gamma factor and the critical synchrotron frequency of the radiating electrons read:
\begin{equation}
\gamma_m=\gamma_{0}\left(\frac{t-t_e}{t_0}\right)^{2 (\alpha_r-1)},
\end{equation}
\begin{equation}
\nu_m=\nu_{m,0}\left(\frac{t-t_e}{t_0}\right)^{(10\alpha_r-s\alpha_r-10)/2}.
\end{equation}

The electron number density within the shocked CSM is given by:
\begin{equation}
n_e=\frac{p-2}{p-1}\frac{(\epsilon_e/\epsilon_B) B_0^2}{8\pi m_ec^2\gamma_{m,0}}\left(\frac{t-t_e}{t_0}\right)^{-s\alpha_r} \rm{cm^{-3}},
\end{equation}
where $m_e$ is the electron mass, and $c$ the speed of light.
The SN progenitor mass loss rate reads:
\begin{equation}
\dot{M}=\frac{8\pi n_e m_pr_0^2v_w}{\eta}\left(\frac{t-t_e}{t_0}\right)^{\alpha_r(2-s)},
\end{equation}
where $m_p$ is the proton mass, $v_w$ is the wind velocity,
while
$\eta$ describes the thickness, $r/\eta$, of the radiating shell at radius $r$. Finally, the ejecta energy reads:
\begin{equation}
E=\frac{4\pi r_0^2B_0^2}{\eta 8\pi \epsilon_{B}}\left(\frac{t-t_e}{t_0}\right)^{5\alpha_r -s \alpha_r - 2}.
\end{equation}
Hereafter, we work under the equipartition hypothesis and set $\epsilon_e=\epsilon_B=0.33$. We note that departures from equipartition would imply $\epsilon_e/\epsilon_B\gtrsim 1$ and likely point to $\epsilon_B\lesssim 0.33$, thus increasing both the shocked electron number density  (and, in turn, the estimated mass-loss rate; see Eqs. (9)-(10)), and the energy budget (Eq. (11)).

As evident from the Equations (1)-(2) and (4)-(8), the observed flux at a given frequency $\nu$ and time $t$ depends on $C_f$, $C_\tau$, $p$, $s$, $\alpha_r$, $\nu_{m,0}$, $\xi$ and $t_e$. Since
$C_f$, $C_\tau$ can be expressed in terms of $r_0$ and $B_0$ (see \cite{skb05} Equations 6-8), the observed flux ultimately depends on $r_0$, $B_0$, $p$, $s$, $\alpha_r$, $\nu_{m,0}$, $\xi$ and $t_e$, which we determine by comparison with the observed data using a $\chi^2$ minimization procedure \citep[see][]{cog14}.
In modeling the radio emission from PTF11qcj, following \citet{cog14}, we set $t_0=10$ days and $\nu_{m,0}\approx 1$\,GHz.
When deriving the mass-loss rate and energy implied by the best-fit results (from Equations (8)-(9)), we also assume a shell thickness of $\eta=10$.

\subsection{Off-axis GRB afterglow model}
In a scenario in which PTF11qcj is powered by a central engine, we may interpret its double-peaked radio light curves as a combination of
radio emission from an uncollimated (non relativistic) SN shock interacting with a dense CSM \citep[as described in][]{cog14}, and that of an off-axis relativistic jet entering our line of sight at late times, as the jet decelerates and spreads sideways \citep{w04}.  The jet dynamics in the relativistic regime is described by the Blandford-Mckee solution \citep{bm76}, and in the late non-relativistic regime by the Sedov-von Neumann--Taylor (SNT) solution \citep{t46}.

Because analytical solutions for the dynamics of a spreading and decelerating relativistic jet cannot fully capture the details (sideways expansion and transition to the non-relativistic regime) of the blast wave evolution, hereafter we use the high-resolution relativistic hydrodynamic simulations by \citet{zm09} and \citet{ehm12} for a jet expanding in a constant density ISM. These two-dimensional simulations include the transition from the relativistic to non-relativistic regime, which is essential to accurately model the GRB outflow at late times \citep{zm09}. The observed flux can be modeled as a function of eight parameters: the isotropic equivalent kinetic energy of the explosion, $E_{\rm iso}$; the circumburst medium number density, $n_{\rm ISM}$; the jet half-opening angle, $\theta_0$; the observer's angle, $\theta_{obs}$; the fraction of internal energy in the shock going into magnetic fields; $\epsilon_{e}$; the fraction of internal energy going into accelerating electrons, $\epsilon_{B}$; the fraction of electrons shock--accelerated in a power--law energy distribution, $\xi_N\sim1$;  and the power-law index of the accelerated electron energy distribution, $p$. These parameters are determined by comparison with the observed fluxes (at each observed frequency and time) using a $\chi^2$ minimization procedure \citep{ehm12}.

\section{Results}
\begin{table}
\begin{center}\begin{footnotesize}
\caption{Best fit parameters for the standard SSA model described in Section 3.1.
Model 0 is the best fit for the first peak in the radio light curves, with fixed $p=3$, $s=2$, and $t_e=55842$.
Models 1 is the best fit for the late-time re-brightening phase.
For Model 1, $B_0$, $s$, and $\xi$ are varied.}
\label{tb:model_results_new_phase2_last6}
\begin{tabular}{lllllllllll}
\hline
\hline
Parameter & Model 0 & Model 1 \\
\hline
$r_0$ (cm)&1.0$\times10^{16}$  &1.0$\times10^{16}$\\
$\xi$&0.24             &0.19\\
$\alpha_r$ &0.81        &0.81\\
$t_e$ &55842            &55842\\
$s$ &2.0               &1.13\\
$B_0$ (G) &6.5         &3.2\\
$p$&3.0               &3.0\\
\hline
$\alpha_B$&-1.0                                &-0.64\\
$\gamma_{m,0}$&7.4                            &11\\
$\alpha_{\gamma}$& -0.38                       &-0.38\\
$n_{e,0}  \, (\rm cm^{-3})$&$1.4\times10^{5}$               &$2.3\times10^{4}$\\
$\alpha_{n_e}$& -1.6                            &-0.91\\
$\dot M_0 \, (\rm M_{\odot}\, yr^{-1})$&9.8$\times10^{-5}$   &$1.7\times10^{-5}$\\
$\alpha_{\dot M}$&0.0                         &0.70\\
$E_{k,0}$ (erg)&7.1$\times10^{48}$                           &1.7$\times10^{48}$\\
$\alpha_{E_k}$&0.42                            &1.1\\
$\chi^2$/dof&1825/90                              &575/35\\
\hline
\end{tabular}
\end{footnotesize}\end{center}\end{table}

\begin{figure*}
    \centering
\includegraphics[width=6.5in,trim={ 1.0cm 12.0cm 0cm 2.5cm},clip]{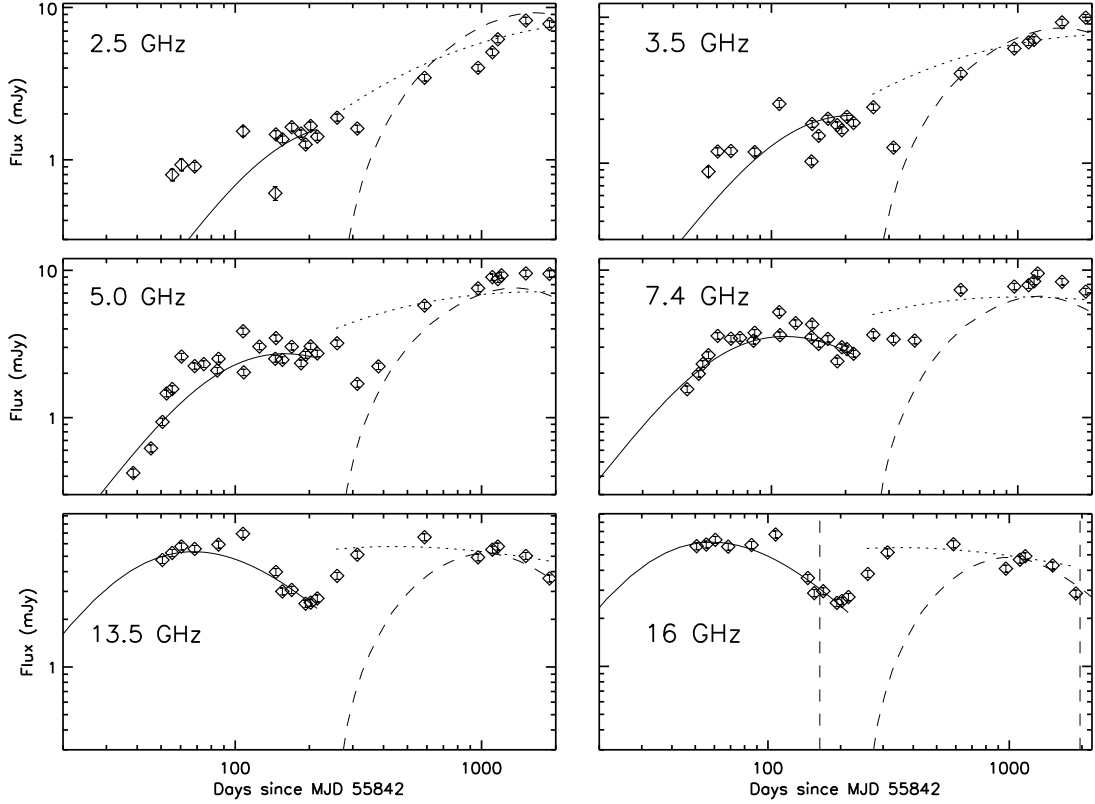}
\caption{Radio light curves of PTF11qcj obtained with the VLA at six different frequencies.
The first radio peak is modeled within a standard SSA model as described in \citet{cog14} (Model 0 in Table~\ref{tb:model_results_new_phase2_last6}; solid curves).
The re-brightening phase is modeled in two different scenarios: (i) within the standard SSA model varying $B_0$, $s$, and $\xi$  (dotted curves; Model 1); and (ii) within an off-axis afterglow model with a constant density ISM and varying $\theta_0$, $E_{\rm iso}$, $n_{\rm ISM}$, and $\theta_{obs}$ (dashed curves). The vertical dashed lines mark the dates of the last Keck and HET spectra on MJD 56006 and 57806 respectively.}
\label{fig:lc_all_models_phase2}
\end{figure*}

\subsection{CSM-interaction radio fits}
In Table~\ref{tb:model_results_new_phase2_last6} and Figure \ref{fig:lc_all_models_phase2} we report fit results for the second radio peak of PTF11qcj within the synchrotron SSA scenario described in Section 3.1 (Model 1). We impose a smooth radial evolution of the SN shock, i.e. we require that  that $r_0$ and $\alpha_r$ remain unchanged with respect to what found during the first radio peak (Model 0), and attempt to model the second radio peak by varying the wind density profile ($s$ and, in turn, $\alpha_B$; see Eq. (5)) and the magnetic field ($B_0$). This is justified by the consideration that a change in these parameters at fixed $r_0$ and $\alpha_r$ effectively corresponds to a change in CSM density as  $F_\nu \propto B^4 \propto n^2_e$ \citep{wsc12}. We also allow for variations in $\xi$. The model was fit to the data obtained after MJD 56429, which is around the time re-brightening occurs, yielding a \textbf{$\chi^{2}/\rm dof=575/35$} (see Table~\ref{tb:model_results_new_phase2_last6}). We note that including the rising data points in the fit results in $B_0\approx1.9$, $s\approx0.87$ and $\xi\approx0.17$, and a \textbf{ $\chi^2/\rm dof=1703/49$}.

\begin{figure*}
    \centering
\includegraphics[width=7.5in,trim={ 1cm 12.0cm 0 1cm},clip]{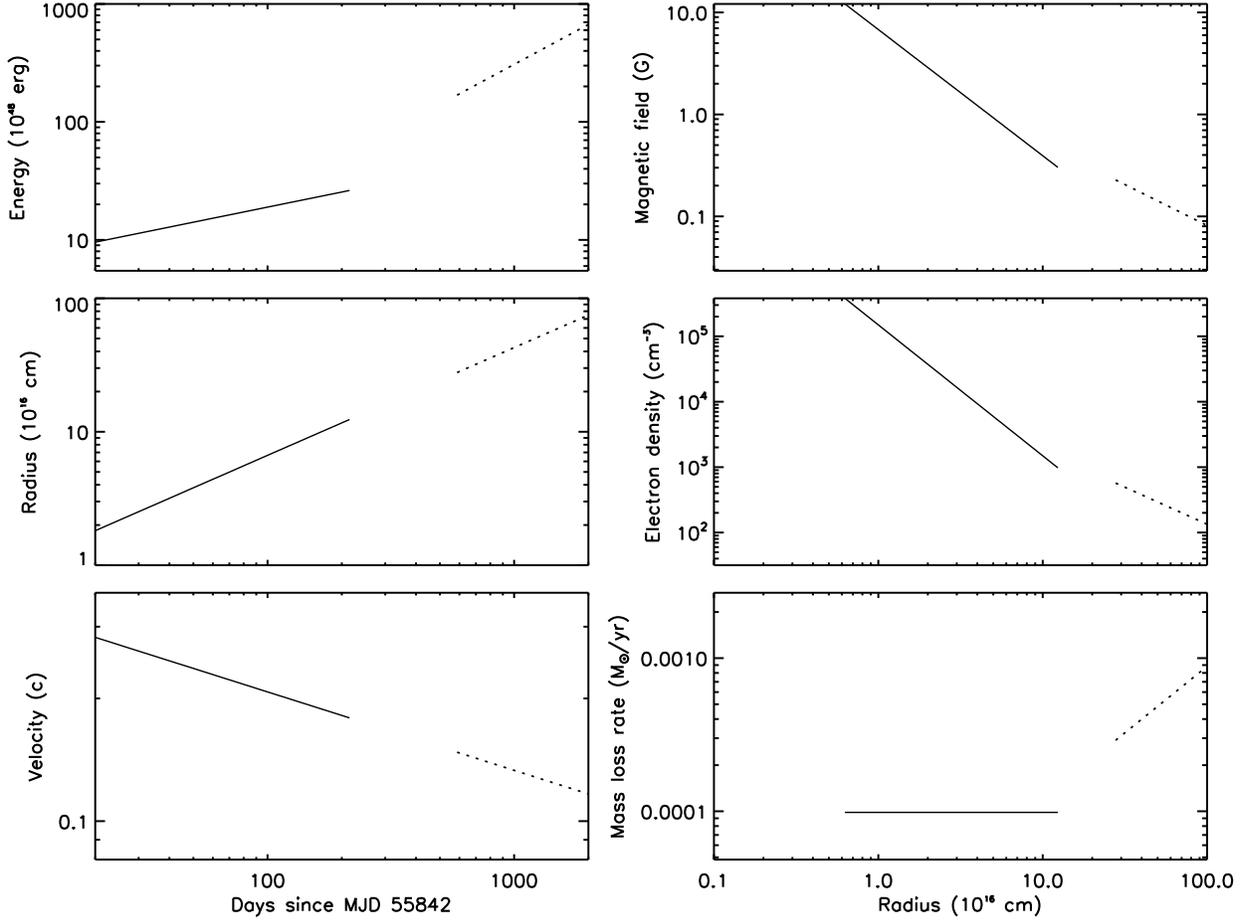}
\caption{Physical parameters derived from Model 0 for phase 1 (solid curves), and from Model 1 (dotted curve) for phase two (re-brightening) of the radio light curve. Left column, from top to bottom: Energy, radius and velocity as a function of time. Right column, from top to bottom: Radial profiles of the magnetic field, electron density, and mass--loss rate.
The best fit models are listed in Table~\ref{tb:model_results_new_phase2_last6}.}
\label{fig:variousparams_all_models_phase2}
\end{figure*}

\begin{figure}
    \centering
\includegraphics[width=3.4in,trim={ 3.0cm 12.5cm 2cm 1cm},clip]{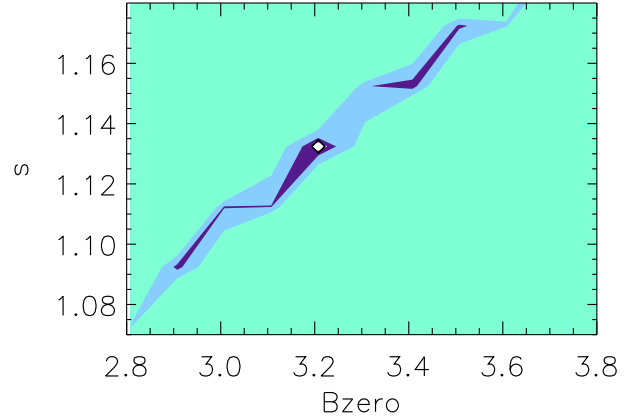}
\label{fig:contours_model1}
\vspace{1mm}
\caption{Best fit results (diamonds)  and confidence intervals for two interesting parameters for Model 1. The white, purple, light blue, and aqua green regions correspond to
confidence intervals of $\lesssim 68\%$, between  $68\%$ and $90\%$, between $90\%$ and $99\%$, and $\gtrsim 99\%$, respectively. See Table \ref{tb:model_results_new_phase2_last6} for more details.}
\label{fig:contours_all}
\end{figure}

The physical parameters derived from Model 1 best fit results are shown in Figure~\ref{fig:variousparams_all_models_phase2}. As evident from this Figure, the best-fit requires an increase in energy and mass-loss rate during the second radio peak. This is similar to what observed in e.g. the CSM-interacting SN\,2003bg \citep{sck06}. In Figure~\ref{fig:contours_all} we show the uncertainties in the best fit results for $s$ and $B_0$, which shows that during the second radio peak a flattening in the CSM profile is also required (compared to the first peak). We note that a flattening in the density profile may be attributed to passage  through a termination shock \citep[e.g.,][]{chev04}, although the simplified analytical model used here does not allow us to properly account for other sources of possible density profile variations such as e.g. clumpiness in the stellar wind.

\vspace{3cm}

\subsection{Off-axis GRB radio fits}
We use the off-axis afterglow model described in Section 3.2 to model the second radio peak in the PTF11qcj light curve. In our fits, $\epsilon_e=\epsilon_B=0.33$, and $p$ are held fixed.
For a fireball expanding in a constant density medium, we find a best fit with an explosion energy of $E_{\rm iso}\approx7\times10^{52}$ erg, $\theta_0\approx0.3$ rad, ISM density $n_{\rm ISM}\approx3\times10^{-5}\,\rm cm^{-3}$, $\theta_{\rm obs}\approx0.6$\,rad, and a $\chi^{2}/\rm dof=1167/36$ (Figure~\ref{fig:lc_all_models_phase2}, dashed curve). The ISM density predicted by this model is rather low compared to typical long GRBs which have $n_{\rm ISM}\gtrsim1\,  \rm cm^{-3}$. Low ISM densities, although peculiar, are not unseen in long GRBs broad-band afterglow modeling \citep{kp00,lbt14,lbm15,alb17}. For completeness, we have also carried out a fit where $n_{\rm ISM}$ is set to the typical long GRB value of $1\,  \rm cm^{-3}$, and $E_{iso}$, $\theta_0$, and $\theta_{\rm obs}$ are allowed to vary. This fit returns $E_{\rm iso}\approx2\times10^{53}$ erg, $\theta_0\approx0.05$ rad, $\theta_{\rm obs}\approx0.17$\,rad, and a $\chi^{2}/\rm dof=4154/37$.

\section{Discussion and conclusion}
We have presented PTF11qcj late--time VLA observations up to $\sim$5 years since optical discovery, and late--time spectroscopic follow--up with the HET at $\approx 5$\,yr post explosion.
The radio luminosity of PTF11qcj is as high as that of the GRB--associated SN\,1998bw.
The radio light curves show a double--peak profile, with the first peak emerging at $\approx$100 days since explosion, and the second at $\approx$2000 days ($\approx 5.5$\,yrs) since explosion. We model the second radio peak (i) with CSM density variations in the standard synchrotron SSA model \citep{skb05}, and (ii) within
an off--axis GRB model  \citep{ehm12}.

We find that density enhancements alone (Model 1) may explain the late--time re-brightening of PTF11qcj.
Radio modeling suggests an enhanced mass-loss rate during the second radio peak.
Even though precursor eruptions have mostly been detected in type IIn SNe, evidence for pre--SN activity was detected in the pre-explosion images of PTF11qcj around May--July 2009 \citep{cog14} hinting at the possibility of such mass--loss episodes being responsible for the enhanced mass--loss rate during the second peak.
Assuming the explosion took place on MJD\,55842, and with $r_0\sim10^{16}$ cm and $\alpha_r\approx0.8$,  the shock radius would have reached $r\approx4\times 10^{17}$\,cm around 1000 days since the explosion (i.e., around the peak of the re-brightening phase).
If material from the pre-SN activity observed in 2009 ($\sim 860$\,d before the explosion) was responsible for the radio re-brightening, the progenitor wind would have traveled at a speed $\sim 24,000\,\rm km s^{-1}$, which is way to high compared to typical stellar winds for stripped-envelope core-collapse SN progenitors ($\sim 1000$\,km\,s$^{-1}$).
The non--detection of H-rich material (to a level distinguishable by our HET/LRS2 spectrum) during the second radio peak also imply that H-rich layers would have been shed well before the 2009 pre-SN activity.

If an off--axis GRB is invoked to explain the late--time radio re-brightening, a
very low $n_{\rm ISM}$ value is required to fit the data. We note that the off--axis afterglow scenario predicts a detectable X--ray flux of $5\times10^{-6}$\,mJy at $\sim$500 days since the explosion, corresponding to $\approx1\times10^{-14}\, \rm erg\,s^{-1}\,cm^{-2}$ at 1 keV - detectable with a 10\,ks observation with \textit{Chandra}. On the other hand, in the SSA model extrapolating the peak radio flux to the X-ray band with a spectral index of $\beta\sim1$ (where $\beta=-(p-1)/2$, with\, $p=3$), gives an expected flux of $\sim4\times10^{-7}$ mJy at 1\,keV (corresponding to $8\times10^{-16} \, \rm erg\,s^{-1}\,cm^{-2}$).
Therefore, we emphasize the importance of observing future events at X--ray frequencies close to the radio peak\footnote{While we have secured a late-time \textit{Chandra} ToO observation of PTF11qcj, execution of our ToO was delayed to 2018 due to scheduling issues.}.

\citet{mmw14} pointed out that in GRB-SNe, the SN carries most of the energy compared to the $\gamma$--ray energy of the jet - an indication that the SN is powered by a central engine.
Since we cannot rule out the off-axis GRB scenario for PTF11qcj, we attempt to estimate the $\gamma$--ray energy of a hypothetical GRB associated with PTF11qcj under the assumption that this was an engine-driven SN.
From the analysis of four GRB-SNe, \citet{l06} finds that the peak spectral energy of GRBs and the peak bolometric luminosity of the underlying supernova are related by
$\nu_{\rm \gamma,peak}=90.2 \,{\rm keV}(L_{\rm SN,peak}/10^{43}\,{\rm erg\,s}^{-1})^{4.97}$.
Considering the peak bolometric luminosity of PTF11qcj $\gtrsim10^{9.3}L_{\odot}$ \citep{cog14}, the peak $\gamma-$ray energy of a hypothetical GRB can be expected to be $\gtrsim 23$ keV.
Then assuming the correlation between the $\nu_{\rm \gamma, peak}$ and $E_{\rm iso}$ as $\nu_{\rm \gamma,peak}=97\, \rm keV(E_{\gamma,iso}/10^{52}\,{\rm erg\,s}^{-1})^{0.49}$ \citep{a06,l06}, we derive $E_{\rm iso}\gtrsim 5\times10^{50}$\,erg may be expected. Incidentally we note that our fit of the second radio peak within the off-axis GRB model implied $E_{\rm iso}\sim 7\times10^{52}$\,erg, so the two results are not in contrast with each other if one assumes a kinetic-energy-to-$\gamma$-ray-energy conversion efficiency of $\gtrsim 1\%$.

We finally note that within the SSA scenario, the expected angular diameter of PTF11qcj would reach the $\sim 1$\,mas level at $\sim 2500$\,d post explosion, or around 6.8\,yr since explosion. A larger angular diameter may be realized if higher ejecta speeds (such as those associated with a GRB jet) would have occurred at any time during the evolution of this explosion. Thus, late-time VLBI observations could potentially probe directly the size of PTF11qcj, and may help distinguish between the standard SSA and off-axis hypothesis.\\
\newline

\acknowledgments A.C. and N.T.P. acknowledge support from the NSF CAREER award \#1455090. J.V. is supported by the GINOP-2.3.2-15-2016-00033 project which is funded by the
Hungarian National Research, Development and Innovation Fund and the European Union.
JCW is supported in part by the Samuel T. and Fern Yanagisawa Regents Professorship. A.G.-Y. is supported by the EU via ERC grant No. 725161, 
the Quantum Universe I-Core program, the ISF, the BSF Transformative program and by a Kimmel award.
The National Radio Astronomy Observatory is a facility of the National Science Foundation operated under cooperative agreement by Associated Universities, Inc.

\end{document}